\documentstyle[12pt]{article}

\newcommand{\resection}[1]{\setcounter{equation}{0}\section{#1}}

\newcommand{\EQ}{\begin{equation}}
\newcommand{\EN}{\end{equation}}
\newcommand{\bea}{\begin{eqnarray}}
\newcommand{\eea}{\end{eqnarray}}
\newcommand{\hs}{\hspace{0.1cm}}

\newcommand{\goto}{\rightarrow}

\begin{document}
\setcounter{page}{0}
\topmargin 0pt
\oddsidemargin 5mm
\renewcommand{\thefootnote}{\arabic{footnote}}
\newpage
\setcounter{page}{0}
\begin{titlepage}
\begin{flushright}
IC/93/143\\
ISAS/EP/93/89
\end{flushright}
\vspace{0.5cm}
\begin{center}
{\large {\bf Mapping between the Sinh-Gordon and Ising Models}}\\
\vspace{1.8cm}
{\large C. Ahn}\\
\vspace{0.5cm}
{\em International Center of Theoretical Physics, \\
Strada Costiera 11, 34014 Trieste, Italy }\\
\vspace{1cm}
{\large G. Delfino, G. Mussardo}\\
\vspace{0.5cm}
{\em International School for Advanced Studies,\\
and \\
Istituto Nazionale di Fisica Nucleare\\
34014 Trieste, Italy}\\

\end{center}
\vspace{1.2cm}

\renewcommand{\thefootnote}{\arabic{footnote}}
\setcounter{footnote}{0}

\begin{abstract}
The $S$-matrix of the Ising Model can be obtained as particular limit
of the roaming trajectories associated to of the $S$-matrix of the
Sinh-Gordon model. Using the form factors of the Sinh-Gordon, we analyse
the correspondence between the operators of the two theories.
\end{abstract}
%\vspace{.3cm}
\end{titlepage}

\newpage

\resection{Introduction}
Given an elastic factorized $S$-matrix of a 2-D system with a mass
scale $M$, we can calculate its ground state energy
$E_0(R)\equiv -\pi\tilde c(MR)/6R$
on an infinite strip of width $R$, by means of the
Thermodynamical Bethe Ansatz (TBA) \cite{Z1,Z2}. For those models where the
$S$-matrix has a well-defined field theory correspondence
\cite{ZZ,Zam1,Musrep}, the scaling function $\tilde c(MR)$ has a smooth
behaviour, monotonically decreasing from the limit value $\tilde c(0)$ (where
it coincides with the effective central charge of the CFT of the ultraviolet
limit) to $\tilde c(\infty)=0$ (which corresponds to massive regime). However,
since the TBA only employs an $S$-matrix without questioning its field
theory interpretation, it can be also used to investigate the finite-size
behaviour of any quantum theory axiomatically defined in terms of a scattering
amplitude, provided it satisfies the usual constraints of unitarity and
crossing symmetry. From this point of view, Al.B.\ Zamolodchikov proposed in
ref.\,\cite{roam} a simple purely elastic scattering theory which under the
TBA analysis reveals a remarkable finite-size behavior. Such theory contains
a single particle bosonic state with mass $M$ and two-particle scattering
amplitude given by
\EQ
S(\beta)=\frac{\sinh\beta-i\cosh\beta_0}{\sinh\beta+i\cosh\beta_0}\,\,,
\label{scattering}
\EN
where $\beta_0$ is a real parameter. $S(\beta)$ has two simple poles in
the unphysical sheet at positions $\beta=-\frac{i\pi}{2}\pm\beta_0$ which
correspond to a resonance particle. The presence of a scale $\beta_0$ for
real values of the rapidities drastically influences the finite-size behaviour
of the model. In fact, solving numerically the TBA equations associated to
the $S$-matrix (\ref{scattering}), for sufficient large values of $\beta_0$,
$\tilde{c}(r)$ develops a ``staircase'' pattern with a series of plateaux
at the discrete values $c\,=\,1-\frac{6}{p(p+1)}$ ($p=3,4,\ldots$) which
coincide with the central charges of unitary minimal models
${\cal M}_p$ \cite{BPZ,FQS}. Hence the Roaming Trajectory Model (RTM) is
a one-parameter family of Renormalization Group flows interpolating between
all the fixed points
${\cal M}_p$: each trajectory starts from the limiting fixed
point ${\cal M}_{\infty}$ and then proceeds on the critical surface through the
hopping ${\cal M}_p \rightarrow {\cal M}_{p-1}$ until it arrives in the
neighborhood of the fixed point ${\cal M}_3$. After this last step, it
develops a finite correlation length and gives rise to the usual massive
infrared behaviour. From the TBA analysis it also follows that the roaming
trajectories spend approximately the same fraction $\beta_0$ of the
Renormalization Group ``time'' $x=\log\,MR/2$ near each fixed point, therefore
making more pronounced the multiple crossover phenomena for large values of
$\beta_0$. Although a local interpretation of the RTM has been pursued in
terms of conformal perturbation of the models ${\cal M}_p$ visiting along the
trajectories \cite{Lassig}, it is worth to consider the RTM as special
analytic continuation of the Sinh-Gordon model in such a way to take advantage
of the recent exact solution of this model \cite{FMS,KM}. Purpose of this
letter is to show, as simplest application of this idea, how to relate the
operator content of the Sinh-Gordon model to that of the Ising model which is
the first jump in the staircase series.

\resection{The Sinh-Gordon model}

\subsection{Main features}

The Sinh-Gordon Model (SGM) is a 2-D Affine Toda Field Theories
\cite{Toda} with one bosonic field $\phi(x)$ and bare action given by
\EQ
A=\int d^2x\,\left[\frac{1}{2}(\partial_\mu\phi)^2-\frac{M_0^2}{g^2}\cosh
g\phi(x)\right]\hs\hs\hs.
\label{action}
\EN
The integrability of the model permits the determination of the factorizable
elastic $S$-matrix which is given by \cite{AFZ}
\EQ
S(\beta,B)=\frac{\sinh\beta-i\sin\frac{\pi B}{2}}{\sinh\beta+i\sin\frac{\pi
B}{2}}
\,\,,
\label{SSG}
\EN
where $B(g)=\frac{2g^2}{8\pi+g^2}$. For real values of the coupling constant
$g$, the $S$-matrix has no poles in the physical sheet and consequently no
bound states, but on the contrary it presents two zeroes at the crossing
symmetric positions $i\pi B/2$ and $i\pi(2-B)/2$. It is easy to see that in the
analytical continuation $B\goto 1\pm\frac{2i}{\pi}\beta_0$ the zeros move
along a direction parallel to the real $\beta$-axis and the $S$-matrix
(\ref{SSG}) exactly coincides with the scattering amplitude of the RTM
\cite{roam}. This observation becomes useful in the light of the fact that
the SGM has been recently solved by computing the matrix elements
of local operators.

\subsection{Form Factors}

A complete knowledge of a QFT is encoded into the matrix elements of
local operators ${\cal O}_k$ on the asymptotic states, the so-called
Form Factors (FF) \cite{Smirnov}
\EQ
F_n^{k}(\beta_1,\ldots,\beta_n)\,=\,<0|{\cal O}_k(0)|\beta_1,\ldots,\beta_n>
\,\,\, .
\EN
In the case of the SGM at real coupling constant, the FF of local scalar
operators have been determined in \cite{FMS,KM}. We briefly recall their main
properties, referring the reader to the original references for their detailed
discussion. They can be parameterized as
\EQ
F_n^{k}(\beta_1,\ldots,\beta_n)\,=\, H_n^k\, Q_n^k(x_1,\ldots,x_n)\,
\prod_{i<j} \frac{F_{\rm min}(\beta_{ij})}{(x_i+x_j)}
\,\,\, ,
\label{para}
\EN
where $x_i\equiv e^{\beta_i}$ and $\beta_{ij}=\beta_i-\beta_j$.
$F_{\rm min}(\beta)$ is an analytic function given by
\begin{eqnarray}
&& F_{\rm min}(\beta,B)\,=\,{\cal N}(B)\,\,\Xi(\beta,B) \nonumber\\
&& \Xi(\beta,B)\,=\,
\exp\,\left[
8\int_0^{\infty} \frac{dx}{x} \frac{\sinh\left(\frac{x B}{4}\right)
\sinh\left(\frac{x}{2}(1-\frac{B}{2})\right) \,\sinh\frac{x}{2}}{\sinh^2 x}
\sin^2\left(\frac{x\hat\beta}{2\pi}\right)\right] \\
&& {\cal N}(B) \,=\,\exp\left[-4\int_0^{\infty}
\frac{dx}{x} \frac{\sinh\left(\frac{x B}{4}\right)
\sinh\left(\frac{x}{2}(1-\frac{B}{2})\right) \,\sinh\frac{x}{2}}{\sinh^2 x}
\right] \nonumber
\end{eqnarray}
($\hat{\beta}=i\pi-\beta$). $F_{\rm min}(\beta,B)$ has a simple zero at
the threshold $\beta=0$ and no poles in the physical strip
$0\leq {\rm Im}\,\beta \leq \pi$, with an asymptotic behaviour
$\lim_{\beta\goto\infty}F_{\rm min}(\beta,B)=1$. In eq.\,(\ref{para}) $H_n^k$
are normalization constants which depend on the operator one is considering.
The functions $Q_n^k(x_1,\dots,x_n)$ are symmetric polynomials in the
variables $x_i$, solutions of the recursion equations which link
the $n$-particle and the $(n+2)$-particle form factors
\EQ
-i\lim_{\tilde\beta \rightarrow \beta}
(\tilde\beta - \beta)
F_{n+2}^k(\tilde\beta+i\pi,\beta,\beta_1,\beta_2,\ldots,\beta_n)=
\left(1-\prod_{i=1}^n S(\beta-\beta_i,B)\right)\,
F_n^k(\beta_1,\ldots,\beta_n)  . \label{rec}
\EN
For FF of spinless operators, the total degree of $Q_n^k$ is equal to
$n(n-1)/2$ whereas their partial degree in each variable $x_i$ depends on
the operator ${\cal O}_k$ which is considered. It was shown in
ref.\,\cite{KM} that a general solution for the $Q_n^k$ can be written in
terms of the so-called {\em elementary solutions} ${\cal Q}_n(p)$ given
by\footnote{We have suppressed the dependence of ${\cal Q}(p)$ from the
variables $x_i$.}
\EQ
{\cal Q}_n(p)\,=\,{\rm det}\,M_{ij}(p)\,\,\, ,
\label{elementary}
\EN
where $M_{ij}(p)$ is an $(n-1)\times (n-1)$ matrix with entries
$ M_{ij}(p)\,=\,\sigma_{2i-j}\,[i-j+p]$ ($\sigma_l$ are the elementary
symmetric polynomials \cite{Macdon} and $p$ an arbitrary integer).

\subsubsection{Form Factors of $\phi(x)$ and $\Theta(x)$}

Important operators of the SGM are the elementary field $\phi(x)$ and the
trace of the stress-energy tensor $\Theta(x)$. They are odd and even operators
respectively under the $Z_2$ symmetry of the model with normalizations given by
$<0 \mid \phi(0)\mid \beta>=1$ and $<\beta\mid\Theta(0)\mid\beta>=2\pi M^2$,
where $M$ is the physical mass. The whole set of FF of the elementary field
$\phi(x)$ is given by
\EQ
F_n^{\phi}(\beta_1,\ldots,\beta_n)\,=\,\left(\frac{4\sin(\pi B/2)}
{{\cal N}(B)}\right)^{(n-1)/2} {\cal Q}_n(0)\,
\prod_{i<j}\frac{F_{\rm min}(\beta_{ij})}{x_i+x_j}\,\,\, .
\label{FFphi}
\EN
They are automatically zero for even $n$ (in agreement with the $Z_2$ parity
of the model) whereas for odd $n$ they vanish asymptotically when
$\beta_i\rightarrow \infty$, as follows from the LSZ reduction formula.
Concerning the FF of $\Theta(x)$, $F_{2n+1}^{\Theta}=0$
whereas $F_{2n}^{\Theta}$ are given by
\EQ
F_{2n}^{\Theta}(\beta_1,\ldots,\beta_{2n})\,=\,\frac{2\pi M^2}
{{\cal N}(B)}\,\left(\frac{4\sin(\pi B/2)}
{{\cal N}(B)}\right)^{n-1} {\cal Q}_{2n}(1)\,
\prod_{i<j}\frac{F_{\rm min}(\beta_{ij})}{x_i+x_j}\,\,\, ,
\label{FFTheta}
\EN
and they go to a constant when $\beta_i\rightarrow \infty$

\subsubsection{Kernel Solutions}

The general structure of the FF of the SGM is that of a sequence of finite
linear spaces whose dimensions grow linearly as $n$ increasing the number
$2n-1$ or $2n$ of external particles. In fact, at each level of
the recursive process the space of the FF is enlarged by including the
kernel solutions of the recursive equation (\ref{rec}), i.e.
$Q_{n}(-x,x,x_1,\ldots,x_{n-2})\,=\,0$. With the constraint that the total
order of the polynomials is $\frac{n(n-1)}{2}$, the kernel is unique and
given by $\Sigma_n(x_1,\ldots,x_n)\,=\,{\rm det}\, \sigma_{2i-j}$. This
solution gives rise to the half-infinite chain under the recursive pinching
$x_1=-x_2=x$
\EQ
\ldots\,\,\rightarrow \,Q_{n+4}^{(n)}\,\rightarrow\,Q_{n+2}^{(n)}\,
\rightarrow \,Q_{n}^{(n)}\,=\Sigma_n\,\rightarrow\,0
\EN
and therefore the whole space of FF presents the foliation
structure\footnote{This is the structure for FF of odd
operators. Analogous structure arises for the FF of even operators.}
\EQ
\begin{array}{ccccccccccccccc}
\ldots & \goto & Q_{n+4}^{(1)} & \goto & Q_{n+2}^{(1)} & \goto & Q_{n}^{(1)} &
\goto & Q_{n-2}^{(1)} &\goto & \ldots & \goto & Q_{3}^{(1)} &\goto & 1 \\
\ldots & \goto & Q_{n+4}^{(3)} & \goto & Q_{n+2}^{(3)} & \goto & Q_{n}^{(3)} &
\goto & Q_{n-2}^{(3)} &\goto & \ldots & \goto & \Sigma_3 & & \\
 & & & . & & .& & . & & . & & . &  & &  \\
 & & & . & & .& & . & & . & & . &  & &  \\
\ldots & \goto & Q_{n+4}^{(n-2)} & \goto & Q_{n+2}^{(n-2)} & \goto
& Q_{n}^{(n-2)} & \goto & \Sigma_{n-2} & &  & & & &  \\
\ldots & \goto & Q_{n+4}^{(n)} & \goto & Q_{n+2}^{(n)} & \goto
& \Sigma_n & & & &  & & & &  \\
\ldots & \goto & Q_{n+4}^{(n+2)} & \goto & \Sigma_{n+2} &
& & & & &  & & & &
\end{array}
\label{chain}
\EN
The explicit expressions of such solutions can be found by determining
the linear combination of ${\cal Q}_n(k)$ which reduces to $\Sigma_n$
at the level $n$.

\resection{Violation of the $c$-theorem sum rule in the RTM}

Since the RTM may be seen as the SGM at $B=1\pm \frac{2i}{\pi}\beta_0$,
it is natural to study the behaviour of the FF of the latter model
under this analytic continuation. As we show, the presence of a scale
$\beta_0$ in the rapidity axes may induce a non-uniform convergence in
series expansions obtained in the original Sinh-Gordon model. Consider for
instance the total variation of the central charge
$\Delta c=c_{\rm uv}-c_{\rm ir}$ going from the short to the large distances.
For both the SGM and the RTM, $\Delta c=1$. Let us try to express it as a
sum-rule fulfilled by the two-point function of the trace $\Theta(x)$
\cite{Zamcth,Cardy}
\EQ
\Delta c \,=\, \frac{3}{4\pi}\,
\int r^2\,<\Theta(r)\Theta(0)>\,d^2r\,=\,
\sum_{n=1}^{\infty} \Delta c^{(2n)}\,\,\, ,
\label{variation}
\EN
where $\Delta c^{(2n)}$ is the contribution to the variation of the central
charge coming from the $2n$-intermediate states. In the original SGM with
real coupling constant, the convergence of the series to the value
$\Delta c=1$ is extremely fast and almost saturated by the two-particle
contribution $\Delta c^{(2)}$ \cite{FMS}. This has to be expected, given the
massive behaviour of the model and the threshold suppression phenomena analyzed
in \cite{Camus}. Similar behaviour has been also observed in supersymmetric
models \cite{Ahn}. However, in the RTM the situation is drastically different.
Consider initially the two-particle contribution to the $c$-theorem sum rule
\EQ
\Delta c^{(2)}(\beta_0)=\frac{3}{2}\int_0^{\infty}d\beta
\,\frac{|\Xi(2\beta,\beta_0)|^2}{\cosh^4\beta}\,\,\, .
\label{320}
\EN
The plot of such a quantity (fig.\,1) shows that $\Delta c^{(2)}(\beta_0)$
monotonically decreases from the value very close to $1$ at $\beta_0=0$
(corresponding to the Sinh-Gordon self-dual point) to $1/2$ for
$\beta_0\goto\infty$. The asymptotic value $1/2$ can be easily
obtained analytically by noticing that
\begin{eqnarray}
&& \Xi(\beta,\beta_0)\,=\,\sinh\frac{\beta}{2}\, h(\beta,\beta_0) \,\, ,\\
&& h(\beta,\beta_0)\simeq -i\left\{
\begin{array}{ll}
\exp\left(-\frac{\beta-\beta_0}{2}\right) & \beta>\beta_0\\
1 & \beta<\beta_0
\end{array}
\right.
\,\,\, ,\nonumber
\end{eqnarray}
and therefore for $\beta_0\goto\infty$ the integral (\ref{320})
simply reduces to
\EQ
\Delta c^{(2)}(|\beta_0|\goto\infty)\,=\,
\frac{3}{2}\int_0^{\infty}d\beta\,\frac{\sinh^2\beta}{\cosh^4\beta}\,=\,
\frac{1}{2}\,\,\, .
\label{319a}
\EN
Concerning the higher particles contributions $\Delta c^{(2n)}$, all of them
vanish in the limit $\beta_0\goto\infty$. In fact, the $2n$-particle FF
entering the formula (\ref{variation}) for $\Delta c^{(2n)}$ is given by
eq.(\ref{FFTheta}) and after the analytic continuation they may be written as
\EQ
F_{2n}(\beta_1,\ldots,\beta_n)\,=\,
2\pi m^2g_{2n}(\beta_0)\,{\cal Q}_{2n}(1)\,
\prod_{i<j}\frac{\sinh\frac{\beta_{ij}}{2}\,
h(\beta_{ij},\beta_0)}
{x_i+x_j}\,\,,
\EN
where $g_{2n}(\beta_0)=(4\cosh\beta_0)^{n-1}{\cal N}^{2n(n-1)}
(\beta_0)$.
Analogously to the two-particle case, the $\beta_0$-dependence coming from
$h(\beta_{ij})$ is strongly suppressed in the integration over rapidities
and the asymptotic behaviour in $\beta_0$ of $\Delta c^{(2n)}$
is only determined by the exponential factors contained in $g_{2n}$ and
${\cal Q}_{2n}(1)$. In the large $\beta_0$ limit,
${\cal N}(\beta_0) \sim \exp\left(-\frac{|\beta_0|}{2}\right)$ and then
$g_{2n}(\beta_0) \sim \exp\left\{-\left(n-1\right)^2|\beta_0|
\right\}$. On the other hand, for $\beta_0\goto\infty$
${\cal Q}_{2n}(1) \sim \exp\left\{(n-1)(n-2)|\beta_0|\right\}
{\cal P}(x_i)$ where ${\cal P}(x_i)$ is a symmetric
polynomial. So, for $n>1$ $\Delta c^{(2n)}(|\beta_0|\goto\infty)\goto 0$
as $\exp\left(-(n-1)\beta_0\right)$. Therefore the result of the series
(\ref{variation}) is $\Delta c=1/2$ instead of $\Delta c=1$, i.e.
a violation of the $c$-theorem sum rule.

Although striking, the non-uniform convergence of the series has a natural
interpretation once the nontrivial interplay between the two scales
$\beta$ and $\beta_0$ of the problem is correctly taken into account. In fact,
since the $n$-particle contribution in (\ref{variation}) behaves as
$e^{-n(Mr)}$, given any length scale $r$ there is always an integer $N_{r}$
such that the states with a number of particles $n\geq N_{r}$ give a
negligible contribution to the series (\ref{variation}). This means that any
partial sum $\Delta c_{N}\equiv\sum\limits_{m=1}^{N}\Delta c^{(2m)}$ only
reproduces the variation of the $c$-function from the infrared limit
$r=\infty$ up to a certain scale $r^{(N)}$. In usual situations, when $c(r)$
is a smooth function in the deep ultraviolet region, the first few
$\Delta c^{(2n)}$ are sufficient to give the correct value of $\Delta c$,
with high level of precision. But for the RTM this is not the case.
Consider a scale $r_1$ such that $c(r_1,\beta_0=0)>1/2$ (fig.\,2). According
to the results of the TBA analysis, after the first jump from $0$ to $1/2$, the
function $c(r,\beta_0)$ stays constant at $1/2$ until a value $r_2$
proportional to $e^{-|\beta_0|/2}$ is reached and, only at this point the
second jump takes place. The other jumps occur at
$r_n\sim e^{-|\beta_0|(n-1)/2}$ and for $\beta_0\goto\infty$, they accumulate
to the origin. Truncating the series (\ref{variation}) to any $N$, there is
{\em always} a value $\beta_0^*$ such that
$c(r_1^{(N)},|\beta_0|>|\beta_0^*|)=1/2$, i.e. the point of the first
jump is always ahead of the corresponding length scale $r_1^{(N)}$, however
small $r_1^{(N)}$ may be, and therefore
\EQ
\lim_{N\rightarrow\infty}\,\lim_{|\beta_0|\goto\infty}
\Delta c_N(\beta_0)=\frac{1}{2} \,\,\, .
\label{nonuniform}
\EN

\resection{Collapse of the Sinh-Gordon Model to the Ising Model}

Taking the limit $\beta_0\goto\infty$ (keeping $\beta$ fixed), the $S$-matrix
of the SGM goes to $S=-1$, i.e. to the $S$-matrix of the thermal perturbed
Ising model. Together with (\ref{nonuniform}), these results naturally suggest
that for $\beta_0\goto\infty$ the Hilbert space of the original SGM collapses
to that of the Ising model, spanned in the local sector only by three
independent families of fields, those of identity $\{1\}$, magnetization
$\{\sigma\}$ and energy $\{\epsilon\}$ operators. It is therefore interesting
to find the mapping between the operator content of the two models.

It is easy to see that the elementary field $\phi(x)$ of the SGM is mapped
onto the magnetization operator $\sigma(x)$ of the Ising model. In fact,
analytically continuing the FF (\ref{FFphi}) and taking the limit
$\beta_0\goto\infty$, the $\beta_0$ dependences coming from different terms
of the original expression compensate each other and we obtain the following
finite result
\EQ
F_{2n+1}^{\phi}(\beta_1,\ldots,\beta_{2n+1})\,\goto\,(i)^{n}
\prod_{i<j}^{2n+1}\tanh\frac{\beta_{ij}}{2}\,\,\, .
\label{magnetization}
\EN
These are precisely the FF of the magnetization operator $\sigma(x)$ of the
thermal perturbed Ising model \cite{KW,YZ}. This field belongs to the
interacting sector of the theory and its correlation functions satisfy
non-trivial differential equations \cite{McCoy,BB}. Notice that in this
limit the boundary conditions of the field $\phi$ have been modified:
in the original SGM its FF vanish for large values of $\beta_i$ whereas
in the resulting expression (\ref{magnetization}) they go to a constant.

On the other hand, taking the limit $\beta_0\goto\infty$ for the analytic
continuation of the FF of $\Theta$ (\ref{FFTheta}), all of them vanish but
$F_2=2\pi m^2\sinh\beta/2$. Hence the operator $\Theta(x)$ of the original
SGM is mapped onto the energy operator $\epsilon(x)$ of the Ising model. This
is a free field (a result which is manifest by the absence of higher FF) and
its correlators can be easily expressed in terms of Bessel functions. Also
in this case the boundary condition of the field $\Theta$ has been changed,
since originally $F_2^{\Theta}$ goes to a constant for large values
of $\beta_i$ whereas after taking the limit $\beta_0\goto\infty$ it
diverges at infinity.

It is also interesting to analyze the behaviour for $\beta_0\goto\infty$
of the kernel solutions. In this limit the recursive equations (\ref{rec})
become
\EQ
\begin{array}{lll}
Q_{n+2}(-x,x,x_1,\ldots,x_n)&=-x^{n+1}\sigma_n\,Q_n(x_1,\ldots,x_n)
& n={\rm odd}\\
Q_{n+2}(-x,x,x_1,\ldots,x_n) & = 0 & n={\rm even}
\end{array}
\label{Isingchain}
\EN
The kernel solutions of the $Z_2$ even operators of the original SGM are
therefore mapped onto the free sectors of the Ising model, i.e. those given
by the identity and energy operators. Indeed, their FF are different from zero
only at a given level $n$ in the number of external particles (where they
coincide with $\Sigma_n$ defined in sec. 2.2.2) and, due to the second
equation in (\ref{Isingchain}), they decouple from the rest of the recursive
chain. Correlators of the operators defined by such FF can be also expressed in
terms of Bessel functions.

Such a decoupling in the recursive chain does not occur, on the contrary, for
the kernel solutions of the odd operators of the original SGM. Their
explicit expressions may be written as determinants of minors of the matrix
$\Sigma_n$. In fact, consider the half-infinite chain of FF $Q^{(n)}_{n+2m}$
($n$ odd and $m=1,2,\ldots$) satisfying the first equation in
(\ref{Isingchain}), with the initial condition
\EQ
Q^{(n)}_{n+2}\,=\,-x^{n+1}\sigma_n\,\Sigma_n\,\,\, .
\EN
It is easy to see that\footnote{We denote by $[A]_{(a,b)}$ the determinant
of the matrix obtained by $A$ eliminating its $a$ row and $b$ column.}
$Q^{(n)}_{n+2}=[\Sigma_{n+2}]_{(\frac{n+1}{2},n-1)}$ and in general
\EQ
Q^{(n)}_{n+2m}\,=\,\left[\left[...[\Sigma_{n+2m}]_
{\left(\frac{n+2m-1}{2},n+2m-1\right)}...\right]_
{\left(\frac{n+3}{2},n+3\right)}\right]_{\left(\frac{n+1}{2},n+1\right)}
\,\,\, .
\EN
Such FF define matrix elements of operators belonging to the
magnetization sector. For instance $Q^{(1)}_n$ defines the FF of the
magnetization operator itself whereas $Q^{(3)}_n$ those of the operator
${\cal O}^{(3)}=(\sigma(x)+1/M^2\partial^2\sigma(x))$ etc.
In general such operators have the distinguishing property that their two-point
correlation function $<{\cal O}^{(n)}(r){\cal O}^{(n)}(0)>$ decreases
at infinity as $\exp[-n Mr]$.

\resection{Conclusions}

The thermal perturbed Ising model is the first model in the staircase
series defined by the RTM. Using the analytic continuation which
links the Sinh-Gordon Model to the RTM, we have seen that in the
limit $\beta_0\goto\infty$ the elementary field of the SGM becomes
the magnetization operator $\sigma$ of the Ising model. This is still
an interacting field with non-trivial form factors. On the other hand,
the field $\Theta$ and other even operators of the SGM are mapped into
the free sectors of the Ising model.

It would be interesting to extend the analysis of this paper to the higher
models of the RTM and to find the correlation functions of the QFT associated
to the corresponding massless Renormalization Group flows.

\vspace{5mm}
\noindent
{\em Acknowledgements}. We are grateful to A. Schwimmer for useful discussions.

\end{document}